\documentclass[prd,superscriptaddress,amsfonts,amssymb,amsmath,showpacs,twocolumn]{revtex4-2}
\usepackage{bm}
\usepackage{amsfonts}
\usepackage{latexsym}
\usepackage{graphicx}
\usepackage{amsmath}
\usepackage{palatino}
\usepackage{mathpazo}
\usepackage{textcomp}
\usepackage{float}
\usepackage{booktabs}
\usepackage{dcolumn}
\usepackage{booktabs}
\usepackage{multirow}
\usepackage{hyperref}
\hypersetup{colorlinks,citecolor=blue}
\usepackage{amsmath}
\usepackage{xcolor}
\usepackage{orcidlink}
\usepackage[caption=false]{subfig}
\usepackage{commath}
\def\jnl@style{\it}
\def\aaref@jnl#1{{\jnl@style#1}}

\def\aaref@jnl#1{{\jnl@style#1}}

\def\aj{\aaref@jnl{AJ}}                   
\def\apj{\aaref@jnl{ApJ}}                 
\def\apjl{\aaref@jnl{ApJ}}                
\def\apjs{\aaref@jnl{ApJS}}               
\def\apss{\aaref@jnl{Ap\&SS}}             
\def\aap{\aaref@jnl{A\&A}}                
\def\aapr{\aaref@jnl{A\&A~Rev.}}          
\def\aaps{\aaref@jnl{A\&AS}}              
\def\mnras{\aaref@jnl{Mon.~Not.~Roy.~Astron.~Soc.}}             
\def\prd{\aaref@jnl{Phys.~Rev.~D}}        
\def\prc{\aaref@jnl{Phys.~Rev.~C}}  
\def\prl{\aaref@jnl{Phys.~Rev.~Lett.}}    
\def\qjras{\aaref@jnl{QJRAS}}             
\def\skytel{\aaref@jnl{S\&T}}             
\def\ssr{\aaref@jnl{Space~Sci.~Rev.}}     
\def\zap{\aaref@jnl{ZAp}}                 
\def\nat{\aaref@jnl{Nature}}              
\def\aplett{\aaref@jnl{Astrophys.~Lett.}} 
\def\apspr{\aaref@jnl{Astrophys.~Space~Phys.~Res.}} 
\def\physrep{\aaref@jnl{Phys.~Rep.}}      
\def\physscr{\aaref@jnl{Phys.~Scr}}       
\def\commat{\aaref@jnl{Comm.~Math.~Phys.}}              
\def\science{\aaref@jnl{Science}}               
\def\cqg{\aaref@jnl{Classical Quant.~Grav.}}            
\def\jpcs{\aaref@jnl{JPCS}}                                     
\def\ijmpd{\aaref@jnl{Int.~J.~Mod.~Phys.~D}}                    
\def\grg{\aaref@jnl{Gen.~Relat.~Gravit.}}               
\def\rpp{\aaref@jnl{Rep.~Prog.~Phys.}}          
\def\npa{\aaref@jnl{Nucl.~Phys.~A}}        
\def\lrr{\aaref@jnl{Living Rev.~Rel.}}                   
\def\jcap{\aaref@jnl{J.~Cosmology Astropart.~Phys.}}    
\def\rmp{\aaref@jnl{Rev.~Mod.~Phys.}}   
\def\epjc{\aaref@jnl{Eur.~Phys.~J.~C}}

\allowdisplaybreaks[1]
\renewcommand{\arraystretch}{1.1}
\begin{document}

\color{black}       

\title{Observational constraints on a logarithmic scalar field dark energy model and black hole mass evolution in the Universe}

\author{Dan Wang\orcidlink{0000-0000-0000-0000}}
\email[Email: ]{sdliuyongedu@163.com}
\affiliation{Department of Mathematics Changzhou College of Information Technology, Changzhou 213164, People's Republic of China}

\author{M. Koussour\orcidlink{0000-0002-4188-0572}}
\email[Email: ]{pr.mouhssine@gmail.com}
\affiliation{Quantum Physics and Magnetism Team, LPMC, Faculty of Science Ben
M'sik,\\
Casablanca Hassan II University,
Morocco.} 

\author{Adnan Malik\orcidlink{0000-0002-6034-0117}}
\email[Email: ]{adnanmalik_chheena@yahoo.com} 
\affiliation{School of Mathematical Sciences, Zhejiang Normal University, \\Jinhua, Zhejiang, China.}
\affiliation{Department of Mathematics, University of Management and Technology,\\ Sialkot Campus, Lahore, Pakistan}

\author{N. Myrzakulov\orcidlink{0000-0001-8691-9939}}
\email[Email: ]{nmyrzakulov@gmail.com}
\affiliation{L. N. Gumilyov Eurasian National University, Astana 010008,
Kazakhstan.}
\affiliation{Ratbay Myrzakulov Eurasian International Centre for Theoretical
Physics, Astana 010009, Kazakhstan.}

\author{G. Mustafa\orcidlink{0000-0003-1409-2009}}
\email[Email: ]{gmustafa3828@zjnu.edu.cn \textcolor{black}{(corresponding author)}}
\affiliation{Department of Physics Zhejiang Normal University Jinhua 321004, People's Republic of China.}
\affiliation{New Uzbekistan University, Mustaqillik Ave. 54, Tashkent 100007, Uzbekistan.}

\date{\today}

\begin{abstract}
We propose a logarithmic parametrization form of energy density for the scalar field dark energy in the framework of the standard theory of gravity, which supports the necessary transition from the decelerated to the accelerated behavior of the Universe. The model under consideration is constrained by available observational data, including cosmic chronometers data-sets (CC), Baryonic Acoustic Oscillation (BAO) data-sets, and Supernovae (SN) data-sets, consisting of only two parameters $\alpha$ and $\beta$. The combined $CC$+$BAO$+$SN$ data-sets yields a transition redshift of $z_{tr}=0.79^{+0.02}_{-0.02}$, where the model exhibits signature-flipping and is consistent with recent observations. For the combined data-sets, the present value of the deceleration parameter is calculated to be $q_{0}=-0.43^{+0.06}_{-0.06}$. Furthermore, the analysis yields constraints on both the parameter density value for matter and the present value of the Hubble parameter, with values of $\Omega_{m0}=0.25849^{+0.00026}_{-0.00025}$ and $H_{0}=67.79_{-0.59}^{+0.59}$ $km/s/Mpc$, respectively, consistent with the results obtained from Planck 2018. Finally, the study investigates how the mass of a black hole evolves over time in a Universe with both matter and dark energy. It reveals that the black hole mass increases initially but stops increasing as dark energy dominates.
\end{abstract}

\maketitle

\section{Introduction}
\label{sec1}

In recent years, cosmological observational data from various probes such as Type Ia supernovae (SN) \cite{Riess, Perlmutter}, BAO \cite{D.J., W.J.}, Cosmic Microwave Background (CMB) \cite{R.R., Z.Y.}, Large Scale Structure (LSS) \cite{T.Koivisto, S.F.}, and recent Planck collaboration \cite{Planck2020} have revealed a startling discovery that has challenged our understanding of the Universe. The expansion of the Universe does not appear to be slowing as expected but rather accelerating at an alarming rate. This unexpected behavior has been attributed to an exotic and poorly understood force known as Dark Energy (DE). DE is a mysterious kind of energy that has a repulsive force that counteracts the attractive force of gravity. This strange phenomenon is supposed to be responsible for the accelerating expansion of the Universe, which was first observed in the late 1990s \cite{Riess, Perlmutter}. Despite numerous efforts, the nature of DE remains a mystery and continues to baffle cosmologists and theorists alike. The discovery of DE has transformed our view of the Universe, and cosmologists have offered various ideas to explain its nature and origin, despite its lack of comprehension. 

Generally, to address the issue at hand, there are two possible approaches that could be pursued. The first involves modifying the energy-momentum tensor, which may be responsible for producing an anti-gravitational force that drives the expansion of the Universe. The second approach involves modifying the geometry component of the Einstein-Hilbert (EH) action, which is equivalent to changing the General Theory of Relativity (GR) \cite{fT1,fQ1,fQ2,fQ3}. In fact, these two approaches aim to provide alternative explanations for the accelerating expansion of the Universe, which is a phenomenon that cannot be accounted for by classical Newtonian mechanics or the original version of GR.

The cosmological constant ($\Lambda$) is by far the most simple and widely accepted candidate. The cosmological constant is a mathematical term developed by Albert Einstein in his GR. It represents a constant energy density that is dispersed uniformly over space-time, and it has an equation of state (EoS) parameter $\omega_{\Lambda}=-1$. This value of $\omega_{\Lambda}$ suggests that the cosmological constant has a negative pressure that produces a repulsive force i.e. $p_{\Lambda}=-\rho_{\Lambda}$, which is assumed to be responsible for the Universe's observed acceleration. Although the cosmological constant has been presented as a candidate for DE, cosmologists are divided on the subject. Some researchers claim that the theoretically expected value of the cosmological constant is much larger than the observed value (fine-tuning problem) \cite{dalal/2001, weinberg/1989}, implying that it may not be the accurate explanation for DE. Others suggest that the cosmological constant is a manifestation of a more complicated and dynamic phenomenon. Therefore, cosmologists have developed a class of models known as dynamical DE models to address these challenges \cite{Ratra,Peebles,Wetterich}. According to these ideas, the DE density is not constant but varies over time, with a rate of change that depends on the evolution of the Universe. These models can better explain observable features of the Universe, such as the acceleration of its expansion, by enabling the DE density to vary dynamically. Dynamical DE models come in various forms, including scalar field models, modified gravity models, and models based on extra dimensions. While each model has its unique features and predictions, they all share the common goal of resolving the difficulties associated with the cosmological constant model. One of the most popular and widely accepted forms of DE is the quintessence scalar field $\phi$ with EoS $\omega>-1$ \cite{Carroll,Y.Fujii}, which is a type of scalar field that serves as a dynamical quantity with a variable density in space-time. Unlike the cosmological constant model, which assumes a constant energy density for DE. Depending on the ratio of its kinetic energy (KE) to potential energy (PE), the quintessence scalar field can be either attractive or repulsive. If the KE is larger than the PE, the scalar field is repulsive and causes the Universe's accelerated expansion. When the PE exceeds the KE, the scalar field attracts and slows the expansion of the Universe. In other terms, if the KE of the scalar field is very small in comparison to PE i.e. $\frac{\overset{.}{\phi }^{2}}{2}<<V\left( \phi \right)$, acceleration in the Universe can be predicted \cite{Q1,Q2,Q3,Q4}. In addition, a number of dynamical DE models exist, including phantom with EoS $\omega<-1$ \cite{Phantom1,Phantom2}, k-essence \cite{T.Chiba,C.Arm.}, Chaplygin gas model \cite{Kamenshchik,M.C.,A.Y.}, tachyon \cite{tachyon}, fermion fields \cite{Myrzakulov} and generalized scalar-fermion fields ($g$-essence) \cite{Yerzhanov}.

Recently, Singh et Nagpal \cite{Singh}, has explored the properties and behavior of DE in the Friedmann-Lemaître-Robertson-Walker (FLRW) cosmology using the EDSFD parametrization. The EDSFD model, which describes the evolution of a scalar field over time, is used to investigate the large-scale evolution of the Universe and the properties of DE. The authors presented analytical and numerical solutions for the EDSFD model and analyze its behavior under different conditions. Pacif et al. \cite{Pacif} used a scalar field source to examine late-time acceleration in the Universe. The authors investigated the observational constraints on the scalar field model and use statefinder diagnostics to analyze the properties of the scalar field. Their results suggest that the scalar field model is consistent with observational data and could provide a viable explanation for the late-time acceleration of the Universe. Moreover, Bairagi \cite{Bairagi} examined the properties of DE models in the context of non-canonical scalar fields in the framework of Einstein-Aether Gravity. The author proposed parametrizations of the DE model that can be used to study the evolution of the Universe and investigate the properties of the non-canonical scalar field. Debnath and Bamba \cite{Debnath} investigated the behavior of DE models in a D-dimensional fractal Universe. The authors considered a non-canonical scalar field in the background of the Universe and propose several parametrizations to model the DE.  Kar et al. \cite{Kar} investigated the relationship between two theoretical frameworks in physics, namely the Dirac-Born-Infeld scalar field DE model and $f(Q)$ gravity. The authors investigated how the coupling between these two models affects the mass accretion of condensed bodies. Sharma, et al. \cite{Sharma} investigated the behavior of the scalar field models of Barrow holographic DE in the context of $f(R,T)$ gravity.

Motivated by the previous discussions, we consider the parametric reconstruction approach to investigate the behavior of the DE model within the framework of GR while considering the presence of the scalar field. This approach involves finding an appropriate parameterization of the cosmological parameters that can accurately represent the evolution of the Universe and its components, such as DE and dark matter, over time \cite{Cunha, Mortsell, Pacif1}. In this work, we have investigated a parametrization of scalar field energy density $\rho_{\phi}$ as a logarithmic function of redshift in flat FLRW space-time (Sec. \ref{sec3} studied the fundamental features of the specified $\rho_{\phi}$), causing a phase transition from early deceleration to present cosmic acceleration. The model parameters are
constrained using the cosmic chronometers (CC), BAO, and SN Ia data-sets. Further, in many cosmological models, the fate of black holes in a Universe filled with DE is still a topic of debate \cite{Babichev,Bamber,Pugliese}. Thus, it is important to have a comprehensive understanding of these astrophysical objects and their behavior throughout the evolution of the Universe \cite{Lima}. This paper aims to investigate the effect of the accretion process of scalar field DE on black holes in a spatially flat FLRW Universe. The paper is presented as follows: The basic equations of GR coupling with the scalar field are presented in Sec. \ref{sec2}. In Sec. \ref{sec3}, we obtain the cosmological solutions using a logarithmic parametrization of the scalar field energy density and then derive the corresponding cosmological parameters. In Sec. \ref{sec4}, we employ the combined $CC$+$BAO$+$SN$ data-sets to obtain the best-fit values of model parameters. Moreover, we investigate the behavior of cosmological parameters for model parameter values constrained by observational data-sets. In Sec. \ref{sec5}, We use statefinder diagnostic to distinguish our scalar field cosmological model from other DE models. Further in Sec. \ref{sec6}, we investigate the evolution of black hole mass. Finally, we discuss and reach a conclusion on our findings in Sec. \ref{sec7}.

Throughout the paper, we have adopted the convention of setting $8\pi G=c=1$.

\section{Basic equations of the model}

\label{sec2}

The gravitational interactions in GR theory are governed by the following
action,%
\begin{equation}
S_{EH}=-\int \frac{1}{2}R\sqrt{-g}d^{4}x+\int L_{m}\sqrt{-g}d^{4}x,
\label{qqm}
\end{equation}%
where $R$ and $L_{m}$ represent the Ricci scalar and the matter Lagrangian
density, respectively. The determinant of the metric is represented by $g$.

The Einstein field equations for the GR theory, derived by varying the
action (\ref{qqm}) with respect to the metric tensor $g_{\mu \nu }$, are
given by%
\begin{equation}
R_{_{\mu \nu }}-\frac{1}{2}g_{_{\mu \nu }}R=T_{\mu \nu },  \label{5}
\end{equation}%
where $T_{\mu \nu }$ is the energy-momentum tensor for the perfect type of
fluid described by%
\begin{equation}
T_{\mu \nu }=\frac{-2}{\sqrt{-g}}\frac{\delta (\sqrt{-g}L_{m})}{\delta
g^{\mu \nu }}.  \label{6}
\end{equation}%

Taking into consideration the spatial isotropy and homogeneity of the
Universe. Here, we presume the flat FLRW metric which represents a
four-dimensional space-time that is both homogeneous, in the sense that it
has the same properties at each and every point in space, and isotropic, in
the sense that it appears the same in all directions. This assumption allows
us to model the Universe's large-scale structure and study the evolution of
its geometry through time,%
\begin{equation}
ds^{2}=dt^{2}-a^{2}(t)[dx^{2}+dy^{2}+dz^{2}],  \label{9}
\end{equation}%
where $a(t)$ is the scale factor used to estimate cosmic expansion at time $%
t $. The Ricci scalar derived corresponding to the above line element is $%
R=-6(\dot{H}+2H^{2})$, where $H=\frac{\overset{.}{a}}{a}$ is the Hubble
parameter. We assumed a flat geometry in our analysis because it is a natural prediction of the inflationary paradigm and is also supported by various cosmological observations, including the CMB measurements \cite{R.R., Z.Y.} and the LSS surveys \cite{T.Koivisto, S.F.}. Further, a flat universe has the advantage of being the simplest and most economical model, as it requires only one more parameter than the base $\Lambda$CDM model.

The energy-momentum tensor for a perfect fluid, which will be assumed here,
is written in a form that takes into account the fluid's energy density,
pressure, and velocity. This tensor is a key notion in GR
that defines how matter and energy are distributed throughout space-time.
When dealing with a perfect fluid, the energy-momentum tensor reduces to a
diagonal form, with each component corresponding to a different physical
quantity of the fluid. This tensor is crucial in predicting the behavior of
astrophysical objects such as stars, galaxies, and even the entire Universe.
The energy-momentum tensor is, $T_{\mu \nu }^{m}=(\rho _{m}+p_{m})u_{\mu
}u_{\nu }-p_{m}g_{\mu \nu }$, where $\rho _{m}$ and $p_{m}$ are respectively
the energy density and pressure of matter.

The Friedmann equations, which explain the dynamics of the Universe in GR
theory, are written as,%
\begin{equation}
3H^{2}=\rho _{m},  \label{fe1}
\end{equation}%
and%
\begin{equation}
2{\dot{H}}+3H^{2}=-p_{m}.  \label{fe2}
\end{equation}

The action of the scalar field is given by a mathematical expression that illustrates how the scalar field interacts with gravity. The scalar field is a hypothetical field that has been postulated to explain numerous physics
phenomena, such as DE, inflation, and the Higgs mechanism. The
action of the scalar field is a fundamental concept in scalar-tensor
theories of gravity, which attempt to generalize Einstein's theory of GR. In
this context, the scalar field plays a crucial role in modifying the
strength of the gravitational force, leading to a rich variety of physical
phenomena. The action of the scalar field (or quintessence model) is usually written as a function
of the scalar field itself, and its derivatives,
\begin{equation}
S_{\phi }=\left[ \frac{1}{2}\partial _{\mu }\phi \partial ^{\mu }\phi
-V\left( \phi \right) \right] \sqrt{-g}d^{4}x,
\end{equation}%
where $\phi $ is the scalar field and $V\left( \phi \right) $ is the scalar
field potential.

Moreover, the action $S_{\phi }$ varies with respect to the scalar field,
leading to the Klein-Gordon equation for metric (\ref{9}) as,
\begin{equation}
\overset{..}{\phi }+3H\overset{.}{\phi }+V^{\prime }\left( \phi \right) =0,
\label{KG}
\end{equation}%
where $V^{\prime }\left( \phi \right) =\frac{dV}{d\phi }$.

Hence, the energy-momentum tensor of the scalar field is derived as,
\begin{equation}
T_{\mu \nu }^{\phi }=(\rho _{\phi }+p_{\phi })u_{\mu }u_{\nu }-p_{\phi
}g_{\mu \nu },
\end{equation}%
where $\rho _{\phi }$ and $p_{\phi }$ are respectively the energy density
and pressure of the scalar field, given by \cite{Barrow1,Barrow2}
\begin{equation}
\rho _{\phi }=\frac{1}{2}\overset{.}{\phi }^{2}+V\left( \phi \right) ,
\label{rho1}
\end{equation}%
and%
\begin{equation}
p_{\phi }=\frac{1}{2}\overset{.}{\phi }^{2}-V\left( \phi \right) .
\label{p1}
\end{equation}

Here, the potential energy $V\left( \phi \right) $ and kinetic energy $\frac{%
\overset{.}{\phi }^{2}}{2}$ are scalar field functions that correspond to
each pair of $(t,x)$ in space-time.

In addition, the scalar curvature coupling, which couples the scalar field
to the Ricci curvature scalar, is commonly used to introduce the coupling
between the scalar field and the gravitational field. This term changes the
gravitational constant, resulting in a change to the Einstein equations. The
coupled action can be expressed as,%
\begin{widetext}
\begin{eqnarray}
S &=&S_{EH}+S_{\phi },  \notag \\
&=&-\int \frac{1}{2\kappa }R\sqrt{-g}d^{4}x+\int L_{m}\sqrt{-g}d^{4}x+
\left[ \frac{1}{2}\partial _{\mu }\phi \partial ^{\mu }\phi -V\left( \phi
\right) \right] \sqrt{-g}d^{4}x.
\end{eqnarray}%
\end{widetext}

Hence, for a general scalar field with the matter as the source, the Friedmann
equations can be rewritten as,%
\begin{equation}
3H^{2}=\rho _{eff},  \label{F1}
\end{equation}%
and%
\begin{equation}
2{\dot{H}}+3H^{2}=-p_{eff},  \label{F2}
\end{equation}%
where $\rho _{eff}=\rho _{m}+\rho _{\phi }$ and $p_{eff}=p_{m}+p_{\phi }$
are the effective, total energy density, and pressure, respectively. Here, we
suppose that the total of the energy and matter included in the Universe is
made up of two types of fluid, one of which corresponds to non-relativistic matter or pressure-less cold dark matter ($p_{m}=0$) and the other a scalar
field, which works as a candidate for dynamical DE (varies with time $t$)
and represents cosmic acceleration in late time.

The continuity equation is written as,
\begin{equation}
\overset{.}{\rho }_{m}+3\rho _{m}H+\overset{.}{\rho }_{\phi }+3\left( \rho
_{\phi }+p_{\phi }\right) H=0.
\end{equation}

Now, we consider that matter and scalar field are both conserved. Hence, the
conservation equations for matter and scalar field are derived as,%
\begin{equation}
\overset{.}{\rho }_{m}+3\rho _{m}H=0,  \label{cm}
\end{equation}%
\begin{equation}
\overset{.}{\rho }_{\phi }+3\left( \rho _{\phi }+p_{\phi }\right) H=0.
\label{cf}
\end{equation}

Solving Eq. (\ref{cm}), we obtain the solution for the matter-energy density 
$\rho _{m}$ as,%
\begin{equation}
\rho _{m}=\rho _{m0}a^{-3},
\end{equation}%
where $\rho _{m0}$ is an arbitrary integration constant and is interpreted
as the current value of the energy density of the matter.

Also, using Eq. (\ref{cf}), we obtain%
\begin{equation}
\overset{.}{\rho }_{\phi }=-3H\left( 1+\omega _{\phi }\right) \rho _{\phi },
\end{equation}%
where $\omega _{\phi }=\frac{p_{\phi }}{\rho _{\phi }}$ represents the EoS
(Equation of State) parameter of the scalar field $\phi $.

Using Eq. (\ref{cf}), the EoS parameter can be calculated as,%
\begin{equation}
\omega _{\phi }=-1-\frac{1}{3H}\left( \frac{\overset{.}{\rho }_{\phi }}{\rho
_{\phi }}\right) ,  \label{EoS}
\end{equation}%
where the dot represents the derivative with respect to cosmic time $t$.

\section{The cosmological model}

\label{sec3}

To solve a system of Eqs. (\ref{F1}) and (\ref{F2}) that comprises just two
independent equations with three unknowns $H$, $\rho _{\phi }$\ and $p_{\phi
}$, we require some extra constraint equations. Investigating models beyond
the cosmological constant is essential since the cosmological constant alone
cannot adequately explain the Universe's accelerating expansion. One method
for investigating these models is to explicitly parametrize the EoS
parameter or the energy density. In the context of DE, the EoS
parameter represents the relationship between pressure and energy density.
The EoS parameter is supposed to be constant across time in the standard
cosmological model, with a value of $-1$ for the cosmological constant.
However, allowing the EoS value to change over time can give insight into
the underlying physics of DE. The Chevallier-Polarski-Linder (CPL)
parametrization is a simple two-parameter model which may capture deviations
from a constant EoS value \cite{CPL1,CPL2}. Much more complicated parametrizations, such as
the Jassal-Bagla-Padmanabhan (JBP) parametrization \cite{JBP1,JBP2}, the logarithmic parametrization \cite{Log1,Log2}, the BA parametrization \cite{BA}, can also be used to
investigate DE scenarios beyond the cosmological constant. Another
method is to parametrize the energy density of DE as a function of
redshift $z$ (or equivalently, cosmic time $t$). This may be accomplished
using a variety of ways, including polynomial expansions and principal
component analysis \cite{Kunz,Mamon1,Mamon2,Das,Singh}. These approaches can shed light on the behavior of DE during various cosmic epochs. Here, we assume that the energy density
for the scalar field is appropriately parametrized as a source of DE in the form,
\begin{equation}
\rho _{\phi }\left( z\right) =\rho _{c0}\log \left( \alpha +\beta z\right) ,
\label{rho_phi}
\end{equation}%
where $\alpha $ and $\beta $\ are constants, and $\rho _{c0}$ is the current
critical density of the Universe. These constants can be estimated from
observational data-sets. The specific choice of a logarithmic energy density for the scalar field is motivated by several reasons. Firstly, logarithmic potentials (or densities) have been explored in the context of scalar fields in various high-energy physics theories and cosmological models. For example, logarithmic potentials can arise naturally in the framework of supersymmetry or string theory, where they can emerge as effective potentials in certain limits or as the result of symmetry breaking mechanisms \cite{Antoniadis,Barrow3,Pallis}. While there is no direct theoretical prediction for the specific logarithmic potential used in our study, it is motivated by the potential relevance and prevalence of logarithmic potentials in the broader context of high-energy physics. Secondly, the logarithmic parametrization allows for a flexible and versatile description of the scalar field's energy density. By introducing logarithmic dependence, the parametrization captures potential deviations from more commonly used functional forms and can account for different behaviors and features of the scalar field \cite{Debnath}. It is important to note that our aim is not to claim a specific high-energy physics theory that predicts the exact form of the logarithmic potential used in Eq. (\ref{rho_phi}). Rather, we adopt this parametrization as a phenomenological approach to investigate the behavior and implications of the scalar field dark energy model. It serves as a tool to explore the compatibility of the model with observational data and to gain insights into the dynamics of DE. In addition, the relationship between redshift $z$ and
scale factor $a\left( t\right) $ is defined by the formula $a\left(
t\right) =\frac{a_{0}}{(1+z)}$, where $a_{0}=1$ is the present value of
scale factor, which leads to the relationship: $\overset{.}{H}=-\left(
1+z\right) H\left( z\right) \frac{dH\left( z\right) }{dz}$. Using this
relationship, we can express the energy density of matter field $\rho _{m}$
as a function of redshift $z$,%
\begin{equation}
\rho _{m}\left( z\right) =\rho _{m0}\left( 1+z\right) ^{3}.  \label{rho_m}
\end{equation}

From Eqs. (\ref{F1}), (\ref{rho_phi}), and (\ref{rho_m}), we obtain%
\begin{equation}
3H^{2}=\rho _{m0}\left( 1+z\right) ^{3}+\rho _{c0}\log \left( \alpha +\beta
z\right) .  \label{Fr}
\end{equation}

Now, we introduce the dimensionless density parameter, which is a measure of
the total density of the Universe relative to the critical density $\rho _{c}
$, i.e. $\Omega =\frac{\rho }{\rho _{c}}$, where $\rho _{c}=3H^{2}$ is
defined as the density necessary for the Universe to have a flat geometry
(i.e., zero curvature). Using Eq. (\ref{Fr}), the dimensionless Hubble
parameter $E\left( z\right) $ in terms of matter density parameter can be
written as,%
\begin{equation}
E^{2}\left( z\right) =\frac{H^{2}\left( z\right) }{H_{0}^{2}}=\Omega
_{m0}\left( 1+z\right) ^{3}+\log \left( \alpha +\beta z\right) ,  \label{H}
\end{equation}%
where $\Omega _{m0}=\frac{\rho _{m0}}{3H_{0}^{2}}$ and $H_{0}$ are the
present values of the matter density and Hubble (i.e. at $z=0$) parameters, respectively.
Eq. (\ref{H}) introduces a logarithmic correction to the DE term,
which modifies the behavior of DE as the Universe evolves. This
modification is a deviation from the standard $\Lambda $CDM model,~which is
the currently accepted standard model of cosmology and is intended to
account for the possibility of new physics beyond the cosmological constant.
The cosmological constant $\Lambda $ is a constant term that does not vary
with time or the expansion of the Universe. However, the logarithmic
correction introduces a dependence on the Hubble parameter, which means that
the DE term can vary as the Universe evolves. For $\Lambda $CDM
model ($\omega _{\phi }=-1$), in the absence of a scalar field, the general
formula for the expansion relation (\ref{H}) is given directly as, $%
E^{2}\left( z\right) =\Omega _{m0}\left( 1+z\right) ^{3}+\Omega _{\Lambda }$%
, where $\Omega _{\Lambda }=\frac{\Lambda }{3H_{0}^{2}}$. Moreover, for $z=0$%
, we can establish an extra constraint on the parameters as $\alpha =\exp
\left( 1-\Omega _{m0}\right) $. This reduces the model's parameters to three
i.e. $H_{0}$, $\Omega _{m0}$, and $\beta $,\ which will be constrained using
the most recent cosmological observational data-sets. For $\Lambda $CDM model,
the extra constraint is given as $\Omega _{m0}+\Omega _{\Lambda }=1$.

The deceleration parameter $q\left( z\right) $ is a dimensionless quantity
that measures the acceleration of the expansion of the Universe. It is
defined as%
\begin{equation}
q\left( z\right) =-\frac{\overset{..}{a}\left( z\right) a\left( z\right) }{%
\overset{.}{a}^{2}\left( z\right) }.
\end{equation}

The deceleration parameter $q\left( z\right) $ determines whether the
Universe's expansion is accelerating or decelerating. If $q<0$, the Universe
is undergoing accelerated expansion, which indicates that the rate of
expansion is accelerating throughout time. This scenario is commonly
associated with the presence of DE, a component that creates
negative pressure and accelerates the expansion of the Universe. On the
other hand, if $q>0$, then the expansion rate of the Universe is decreasing
over time. This means that the Universe is decelerating, which can be caused
by the presence of matter or radiation, both of which create positive
pressure and resist the Universe's expansion. It is interesting to note that
the deceleration parameter $q\left( z\right) $ can also be represented in
terms of the dimensionless Hubble parameter, as shown below,%
\begin{equation}
q(z)=-1+\frac{\left( 1+z\right) }{E\left( z\right) }\frac{dE\left( z\right) }{%
dz}.  \label{q}
\end{equation}

Using Eqs. (\ref{H}) and (\ref{q}), we get%
\begin{equation}
q(z)=-1+\frac{(1+z)\left[ 3\Omega _{m0}(1+z)^{2}+\frac{\beta }{\alpha +\beta
z}\right] }{2\left[ \Omega _{m0}(1+z)^{3}+\log (\alpha +\beta z)\right] }.
\label{qz}
\end{equation}

Using Eq. (\ref{EoS}), the EoS parameter for the scalar field can be written
in terms of redshift $z$ as,%
\begin{equation}
\omega _{\phi }\left( z\right) =-1+\frac{\beta (1+z)}{3(\alpha +\beta z)\log
(\alpha +\beta z)}.
\end{equation}

Hence, the effective EoS parameter for our model is,%
\begin{equation}
\omega _{eff}=\frac{p_{eff}}{\rho _{eff}}=\frac{p_{\phi }}{\rho _{m}+\rho
_{\phi }},
\end{equation}%
where $p_{eff}$ and $\rho _{eff}$ represent the total pressure and energy
density of the Universe, respectively. So we have%
\begin{equation}
\omega _{eff}\left( z\right) =\frac{\beta (1+z)-3(\alpha +\beta z)\log
(\alpha +\beta z)}{3(\alpha +\beta z)\left[ \Omega _{m0}(1+z)^{3}+\log
(\alpha +\beta z)\right] }.
\end{equation}

The density parameters $\Omega _{m}$ and $\Omega _{\phi }$ for matter and
scalar field in terms of $z$ are obtained as,%
\begin{equation}
\Omega _{m}\left( z\right) =\frac{\rho _{m}}{3H^{2}}=\frac{\Omega
_{m0}(1+z)^{3}}{\log (\alpha +\beta z)+\Omega _{m0}(1+z)^{3}},
\end{equation}%
and%
\begin{equation}
\Omega _{\phi }\left( z\right) =\frac{\rho _{\phi }}{3H^{2}}=\frac{\log
(\alpha +\beta z)}{\log (\alpha +\beta z)+\Omega _{m0}(1+z)^{3}},
\end{equation}%
respectively.

From Eqs. (\ref{rho1}), (\ref{p1}), (\ref{F1}) and (\ref{F2}), the
expressions for the potential energy $V\left( \phi \right) $ and kinetic
energy $\frac{\overset{.}{\phi }^{2}}{2}$ of the scalar field are obtained
as,%
\begin{equation}
V\left( \phi \right) =\frac{H_{0}^{2}}{2}\left[ 6\log (\alpha +\beta z)-%
\frac{\beta (1+z)}{\alpha +\beta z}\right] ,
\end{equation}%
and%
\begin{equation}
\frac{\overset{.}{\phi }^{2}}{2}=\frac{\beta H_{0}^{2}(z+1)}{2(\alpha +\beta
z)}.
\end{equation}%
respectively. The model described in Eq. (\ref{H}) is heavily influenced by its parameters $(H_{0},\Omega _{m0},\beta )$, which dictate its behavior and cosmological properties. To better understand the implications of this model, we analyze recent observational data-sets in the next section. Specifically, they aim to constrain the values of key parameters $(H_{0},\Omega _{m0},\beta )$ and examine how this affects the behavior of cosmological parameters. 

\section{Data fittings and numerical results}

\label{sec4}

To examine the validity of this scenario, we employ observational data-sets
from Hubble measurements, BAO, and pantheon SNIa samples integrated with
other SNIa data points. In this section, we'll go through how we use these
data-sets. To adapt the data-sets, we employ Bayesian statistical analysis
and the emcee package \cite{Mackey/2013} to perform a Markov chain Monte
Carlo (MCMC) simulation.

\subsection{Cosmic Chronometer (CC) data-sets}

The $\chi ^{2}$ function is a statistical tool used in cosmology to
determine the best-fit values for the parameters of a given cosmological
model based on observational data. In this case, we are considering Hubble
parameter measurements derived from the differential age (DA) method, which
is also known as CC data-sets. To be more exact, we are using 31 points from the
Refs. \cite{Yu/2018,Moresco/2015,Sharov/2018}. These points show Hubble
parameter observations at various redshifts, which can be employed to
constrain the Universe's expansion history and test various cosmological
scenarios. The $\chi ^{2}$ function is defined as,%
\begin{equation}
\chi _{CC}^{2}=\sum_{i}\frac{\left[ H(\theta ,z_{i})-H_{obs}(z_{i})\right]
^{2}}{\sigma (z_{i})^{2}},
\end{equation}%
where $i$ runs over the $31$ data points, $H(\theta _{s},z_{i})$ denotes the
predicted value of the Hubble parameter at redshift $z_{i}$ for a given set
of cosmological parameters $\theta =(H_{0},\Omega _{m0},\beta )$, $%
H_{obs}(z_{i})$ denotes the observed value of the Hubble parameter at
redshift $z_{i}$, and $\sigma (z_{i})$ denotes the uncertainty in the
measurement of $H_{i}$. The goal is to find the values of $\theta $ that
minimize the value of $\chi _{CC}^{2}$. Typically, an MCMC approach is used,
which explores the parameter space and determines the regions with the
greatest likelihood given the data.

\subsection{Baryon Acoustic Oscillations (BAO) data-sets}

BAOs are a feature of the large-scale structure of the Universe that emerge
from primordial density perturbations in the baryon-photon plasma in the
early Universe. These perturbations produced pressure waves, which
propagated through the plasma until the Universe became transparent to
photons, a \ process known as recombination. At this point, the pressure
waves imprinted a characteristic scale on the distribution of matter, which
can still be seen in the clustering of galaxies today. BAO emerge as peaks
in the cosmic microwave background radiation power spectrum and in the
distribution of galaxies on large angular scales. We can deduce the
characteristic scale of the BAO by measuring the position of these peaks,
which is connected to the sound horizon at recombination and serves as a
standard ruler for cosmic distance measurements. In this regard, we use BAO
data-sets from several surveys, including the Six Degree Field Galaxy Survey
(6dFGS), the Sloan Digital Sky Survey (SDSS), and the LOWZ samples of the
Baryon Oscillation Spectroscopic Survey (BOSS) \cite%
{Blake/2011,Percival/2010,Giostri/2012}. These studies have yielded precise
measurements of the positions of the BAO peaks in the clustering of galaxies
at various redshifts, allowing us to constrain the expansion history of the
Universe and test various cosmological scenarios. This work examines six points of BAO data-sets as well as the cosmology stated below, 
\begin{equation}
d_{A}(z)=\int_{0}^{z}\frac{dy}{H(y)},
\end{equation}%
\begin{equation}
D_{v}(z)=\left[ \frac{d_{A}^{2}(z)z}{H(z)}\right] ^{1/3},
\end{equation}%
where $d_{A}(z)$ is the comoving angular diameter distance, and $D_{v}$ is the
dilation scale. Moreover, the $\chi ^{2}$
function for BAO data-sets is defined as, 
\begin{equation}
\chi _{BAO}^{2}=X^{T}C_{BAO}^{-1}X.
\end{equation}

Here, $X$ depends on the considered survey and $C_{BAO}^{-1}$ is the inverse
covariance matrix \cite{Giostri/2012}.

\subsection{Type Ia Supernova (SN Ia) data-sets}

SN Ia are one of the most significant cosmic probes used to investigate the
nature of DE and the Universe's accelerating expansion. These SN
are assumed to be the consequence of a white dwarf star exploding in a
binary system, and they have a highly distinctive light curve, making them
great "standard candles" for estimating cosmic distances. By comparing the
observed luminosity of SN Ia to their predicted intrinsic luminosity, we
can determine their distance from us and chart the expansion history of the
Universe. This approach has been widely employed in modern cosmology and has
produced significant evidence for the presence of DE, the enigmatic
component responsible for the Universe's accelerating expansion. In this
perspective, the Pantheon collection of SN Ia is an especially valuable
data-sets, with 1048 data points spanning a wide range of redshifts, from $0.01
$ to $2.26$. The Pantheon sample is constructed from the DE Survey
and the Supernova Legacy Survey, among other sources, and has been
extensively calibrated to eliminate systematic errors and increase the
accuracy of distance estimations \cite{Scolnic/2018,Chang/2019}.

The $\chi ^{2}$ function for SN data-sets is defined as,%
\begin{equation}
\chi _{SN}^{2}=\sum_{i,j=1}^{1048}\Delta \mu _{i}\left( C_{SN}^{-1}\right)
_{ij}\Delta \mu _{j},
\end{equation}%
where $\Delta \mu _{i}=\mu _{th}-\mu _{obs}$ represents the difference
between the observational and theoretical distance modulus, and $C_{SN}^{-1}
$ denotes the data's inverse covariance matrix. In addition, we define $\mu
=m_{B}-M_{B}$, where $m_{B}$ represents the measured apparent magnitude at a
certain redshift and $M_{B}$ represents the absolute magnitude. The nuisance
parameters in the previous equation were calculated using a new approach
called as BEAMS with Bias Corrections (BBC) \cite{Kessler/2017}. The
theoretical value is calculated as,%
\begin{equation}
\mu _{th}(z)=5log_{10}\frac{d_{L}(z)}{1Mpc}+25,
\end{equation}%
\begin{equation}
d_{L}(z)=(1+z)\int_{0}^{z}\frac{dy}{H(y,\theta )},
\end{equation}%
where $d_{L}(z)$ represents the luminosity distance.

\subsection{CC+BAO+SN data-sets}

To perform the combined data-sets: CC, BAO, and SN samples, we employ the total
joint $\chi ^{2}$ function as, $\chi _{total}^{2}=\chi _{CC}^{2}+\chi
_{BAO}^{2}+\chi _{SN}^{2}$. The best-fit values of the model parameters can
be estimated by minimizing the corresponding $\chi ^{2}$ value, which is
analogous to the maximum likelihood analysis. By using the aforementioned
combined CC+BAO+SN data-sets, Fig. \ref{CC+BAO+SN} depicts the statistical
findings for the model with $1-\sigma $ and $2-\sigma $ likelihood contours.
Tab. \ref{tab} also corresponds to the values of the parameter space
estimated from the combined data-sets. Figs. \ref{ErrorHubble} and %
\ref{ErrorSNe} depict the error bar plots for the considered model and the $%
\Lambda $CDM or standard cosmological model, with matter density parameter $%
\Omega _{m}^{0}=0.315\pm 0.007$ and $H_{0}=67.4\pm 0.5$ $Km/s/Mpc$ \cite%
{Planck2020}. So, as seen in the figures, our model closely matches the
observed data. In this study, we have combined data-sets from three different observational techniques, namely, CC, BAO, and SN data-sets, to estimate the current value of $H_0$ at $z=0$. Our analysis yielded a value of $H_{0}=67.79_{-0.59}^{+0.59}$ $km/s/Mpc$, with a statistical uncertainty of $\pm0.59km/s/Mpc$. This is in agreement with the recent findings of Aubourg et al. \cite{Aubourg}, who also utilized a combination of BAO measurements, CMB data, and a reanalysis of SN data to constrain cosmological parameters and test DE models. Remarkably, our results align with their outcomes, further supporting the robustness of our analysis. Notably, our approach is model-independent, allowing us to obtain consistent measurements of the present value of the Hubble parameter $H_{0}$ that is fully in line with the results from the Planck-2018 study on the standard cosmological model ($\Lambda$CDM) \cite{Planck2020}. Moreover, the parameterization method has been employed in various DE models to obtain a comparable value for the Hubble parameter $H_{0}$ \cite{Chen1, Chen2, Capozziello}. 

\begin{widetext}

\begin{figure}[h]
\centerline{\includegraphics[scale=0.8]{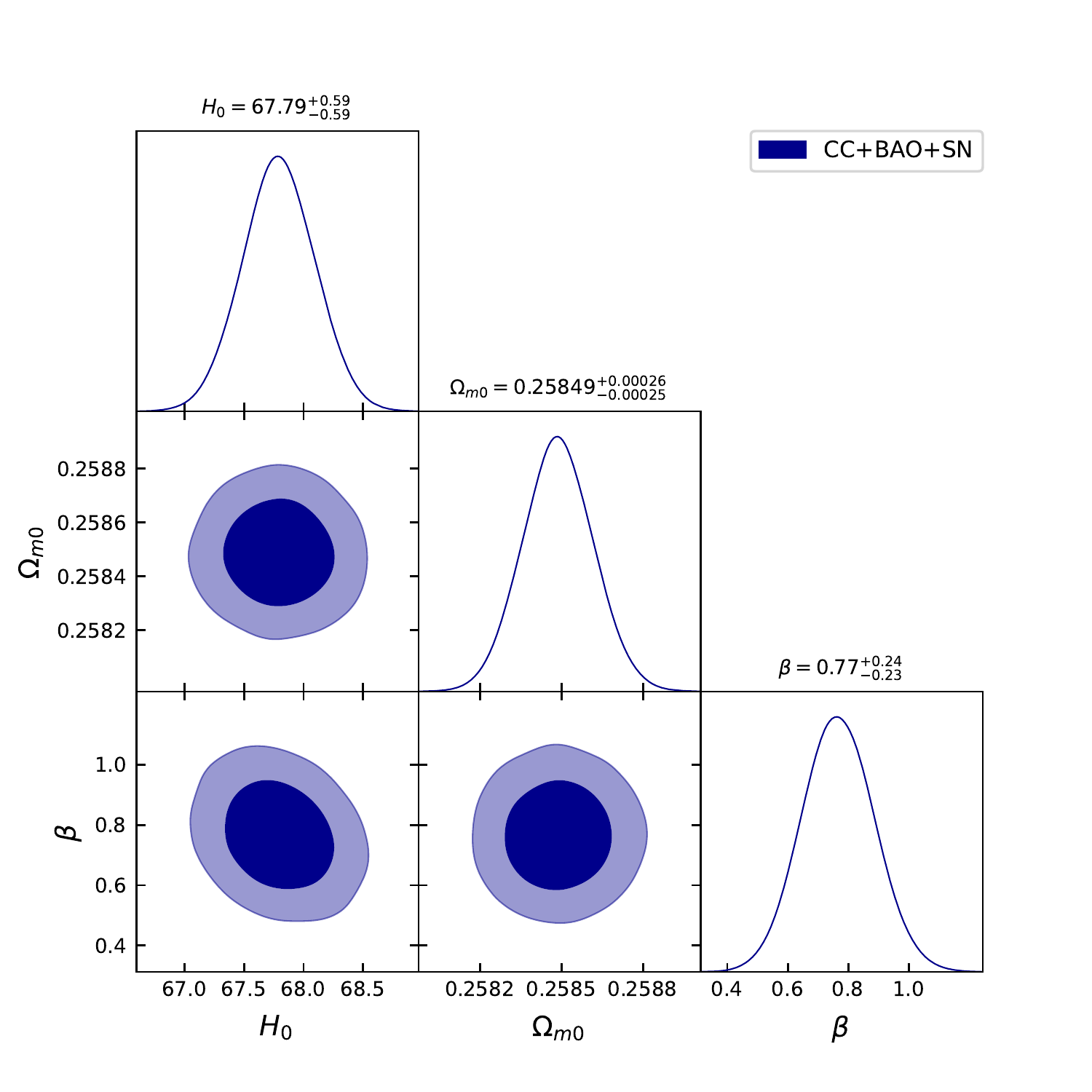}}
\caption{The curves for the $1-\sigma$ and $2-\sigma$ confidence levels can be observed for the model parameters $H_{0}$, $\Omega_{m0}$, and $\beta$, when using the $CC$+$BAO$+$SN$ data-sets.}
\label{CC+BAO+SN}
\end{figure}

\begin{figure}[h]
\centerline{\includegraphics[scale=0.60]{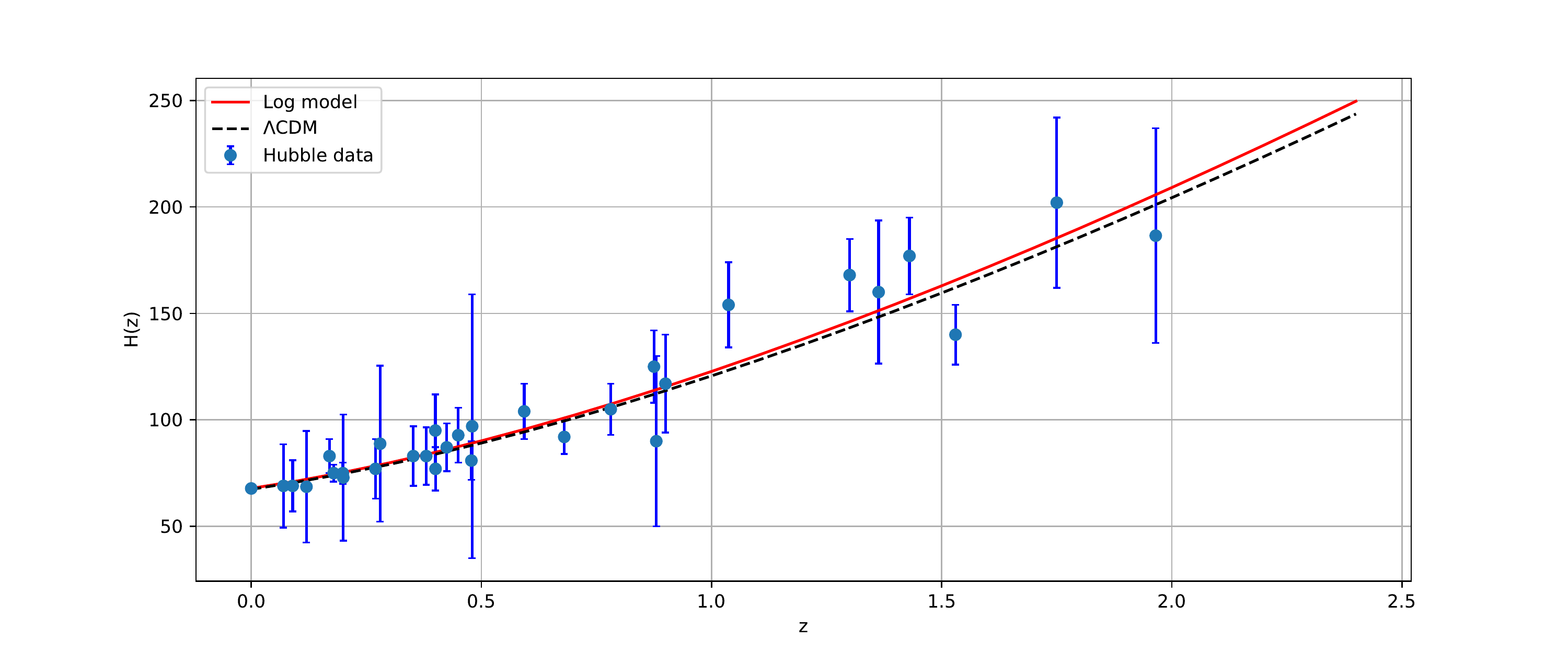}}
\caption{The variation of Hubble parameter $H(z)$ concerning redshift $z$ can be observed through the graph. The black dashed line displays the $\Lambda$CDM model, while the red line represents the logarithmic model curve. The blue dots in the graph depict error bars.}
\label{ErrorHubble}
\end{figure}

\begin{figure}[h]
\centerline{\includegraphics[scale=0.60]{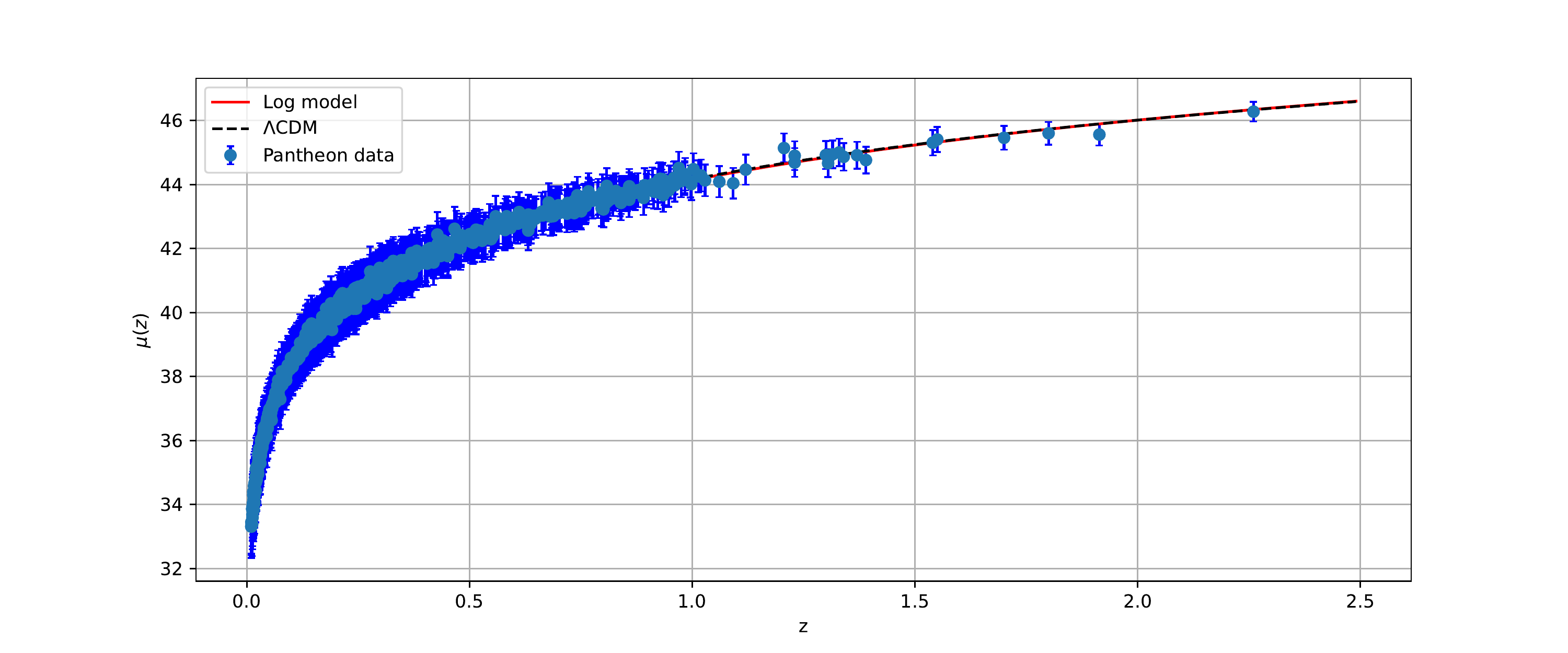}}
\caption{The variation of distance modulus $\mu(z)$ concerning redshift $z$ can be observed through the graph. The black dashed line displays the $\Lambda$CDM model, while the red line represents the logarithmic model curve. The blue dots in the graph depict error bars.}
\label{ErrorSNe}
\end{figure}

\begin{table*}[!htb]
\begin{center}
\renewcommand{\arraystretch}{1.5}
\begin{tabular}{l c c c c c c c c c c}
\hline 
Parameters  & $H_{0}$ & $\Omega _{m0}$ & $\alpha=\exp
\left( 1-\Omega _{m0}\right)$ & $\beta$ & $q_{0}$ & $z_{tr}$ & $\omega _{0}$\\
\hline
$Priors$ & $(60,80)$ & $(0,1)$  & $-$ & $(-10,10)$ & $-$ & $-$ & $-$\\

$CC+BAO+SN$   & $67.79\pm0.59$  & $0.25849^{+0.00026}_{-0.00025}$  & $2.0991^{+0.00054}_{-0.00052}$ & $0.77^{+0.24}_{-0.23}$ & $-0.43^{+0.06}_{-0.06}$ & $0.79^{+0.02}_{-0.02}$ & $-0.84^{+0.06}_{-0.05}$\\
\hline
\end{tabular}
\caption{The Best-fit values for the parameter space can be determined by utilizing the combined $CC$+$BAO$+$SN$ data-sets.}
\label{tab}
\end{center}
\end{table*}
\end{widetext}

\subsection{The model's cosmological parameters and their behavior}

\begin{figure}[h]
\centerline{\includegraphics[scale=0.70]{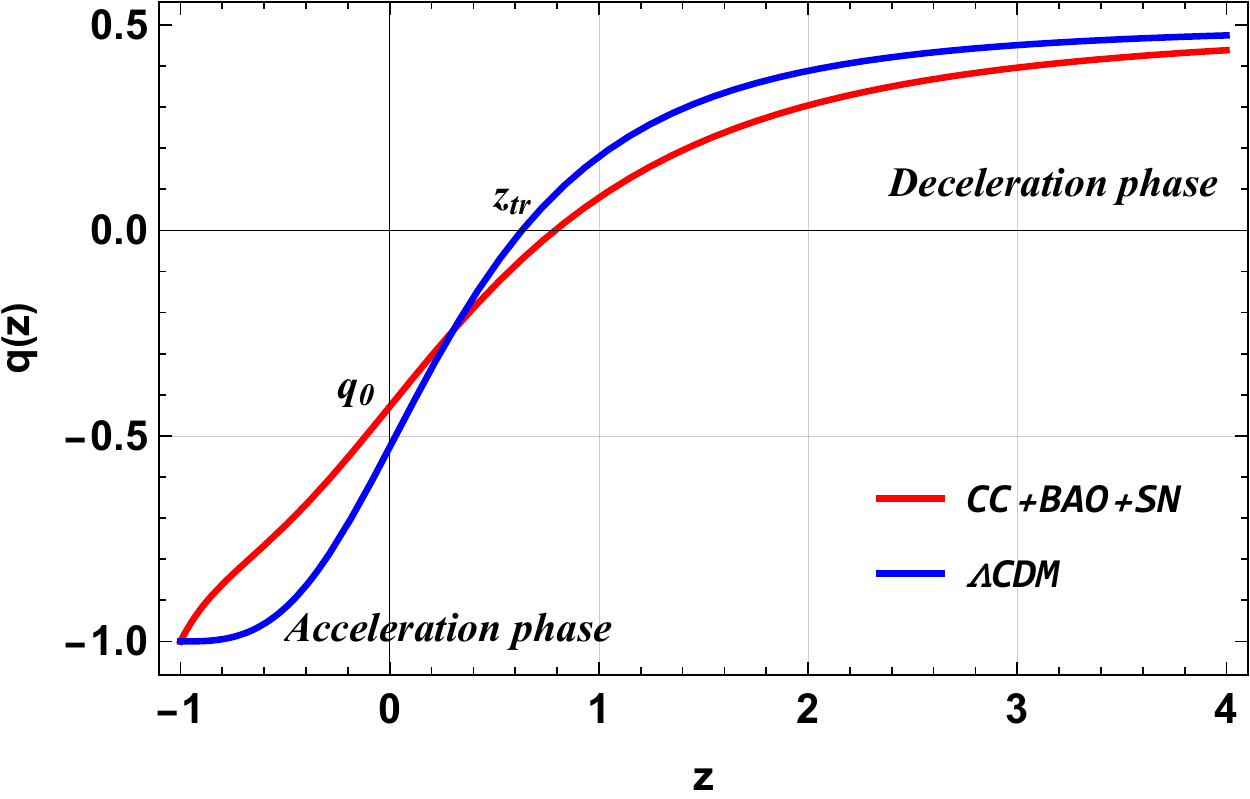}}
\caption{The plot shows the relation between the deceleration parameter ($q$) and redshift ($z$) using the values constrained from the combined $CC$+$BAO$+$SN$ data-sets. The figure also includes a comparison between our model and the $\Lambda$CDM.}
\label{F_q}
\end{figure}

 The deceleration parameter plays a fundamental role in understanding the dynamics and evolution of the Universe. By studying the rate at which the Universe is expanding, we can learn about its past and future behavior. Eq. (\ref{qz}) used to calculate the deceleration parameter contains three parameters that are constrained by observational data. With these numerical data, we can examine the evolution of the deceleration parameter and draw predictions about the Universe's expansion rate. This knowledge can help us better comprehend the fundamental features of the Universe. Furthermore, investigating the deceleration parameter might help us assess the validity of our cosmological model. Whereas the negative value of $q$ represents the accelerating phase, the positive value of $q$ represents the decelerating phase. From Fig. \ref{F_q}, the deceleration parameter $q$ varies with $z$ from positive to negative. This shows a transition from early deceleration $q>0$ to the Universe's current acceleration $q<0$. The transition redshift is $z_{tr}=0.79^{+0.02}_{-0.02}$ corresponding to the combined $CC$+$BAO$+$SN$ data-sets \cite{Farooq, Jesus, Garza}. The present value of the deceleration parameter is $q_{0}=-0.43^{+0.06}_{-0.06}$. Recently, in a study by Capozziello et al. \cite{Capozziello}, an empirical value of the deceleration parameter $q_0$ was determined to be $-0.56$ with a statistical uncertainty of $\pm0.04$. This finding highlights the significance of $q_0$ in understanding the dynamics of the Universe. In addition, several other empirical values of $q_0$ in the vicinity of the results obtained in our analysis can be found in the references provided \cite{Mamon1, Mamon2, Basilakos}.

\begin{figure}[h]
\centerline{\includegraphics[scale=0.70]{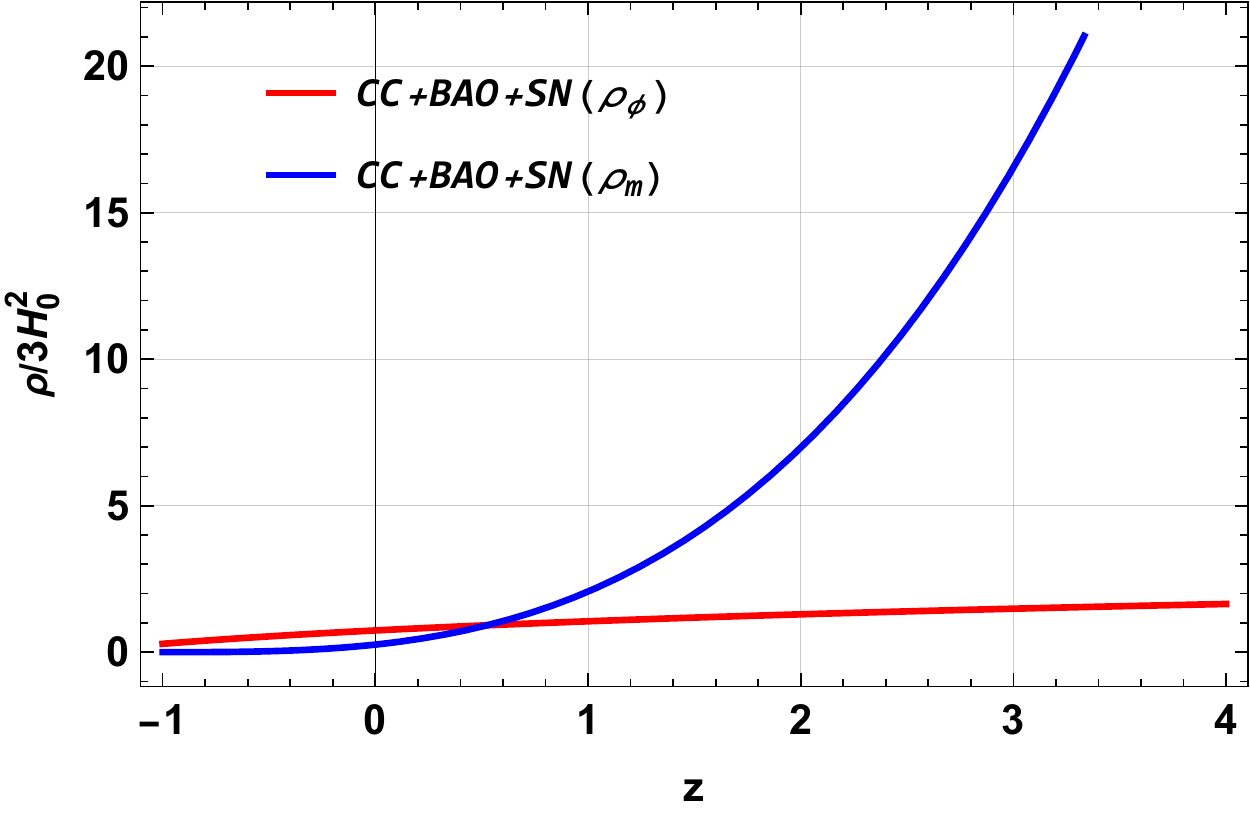}}
\caption{The plot shows the relation between the densities of scalar field DE and matter ($\rho_{\phi}$ and $\rho_{m}$) and redshift ($z$) using the values constrained from the combined $CC$+$BAO$+$SN$ data-sets.}
\label{F_rho}
\end{figure}

 According to Fig. \ref{F_rho}, the densities of matter and scalar field DE decrease as the Universe expands. In late time, the matter density approaches zero, while the scalar field DE density approaches a minimal value. The decrease in the matter and scalar field DE densities as the Universe expands is due to energy conservation in GR. The idea that matter density tends to zero in the late Universe has important consequences for the Universe's future evolution. The Universe will grow progressively dark and cold when there is no matter left to form new stars or galaxies, a condition called the "heat death" of the Universe. This scenario results from the second law of thermodynamics, which states that entropy, or disorder, can constantly rise over time. In addition, since scalar field DE density tends to have a small value means that the Universe can continue to expand at an accelerating rate in the future. This is known as the "big rip" scenario, in which the Universe's expansion grows so rapidly that it pulls apart all matter, including galaxies and stars.

In this context, the EoS parameter is a  good tool for describing the Universe's behavior in terms of its expansion rate. The EoS parameter combines the pressure $(p)$ and energy density $(\rho)$ of the cosmic fluid and is defined as $\omega = p/\rho$. Depending on the nature of the cosmic fluid, the EoS parameter $\omega$ can have different values. For non-relativistic matter, such as dark matter, $\omega = 0$. In addition, $\omega =\frac{1}{3}$ in the case of relativistic matter, such as radiation. The value of the EoS parameter $\omega$ can be used to classify the Universe's decelerating and accelerating behavior. There are three possible eras for a Universe with positive acceleration:
\begin{itemize}
    \item Quintessence era: $-1<\omega<-\frac{1}{3}$,
    \item Phantom era: $\omega<-1$,
    \item Cosmological constant era: $\omega=-1$.
\end{itemize}

\begin{figure}[h]
\centerline{\includegraphics[scale=0.72]{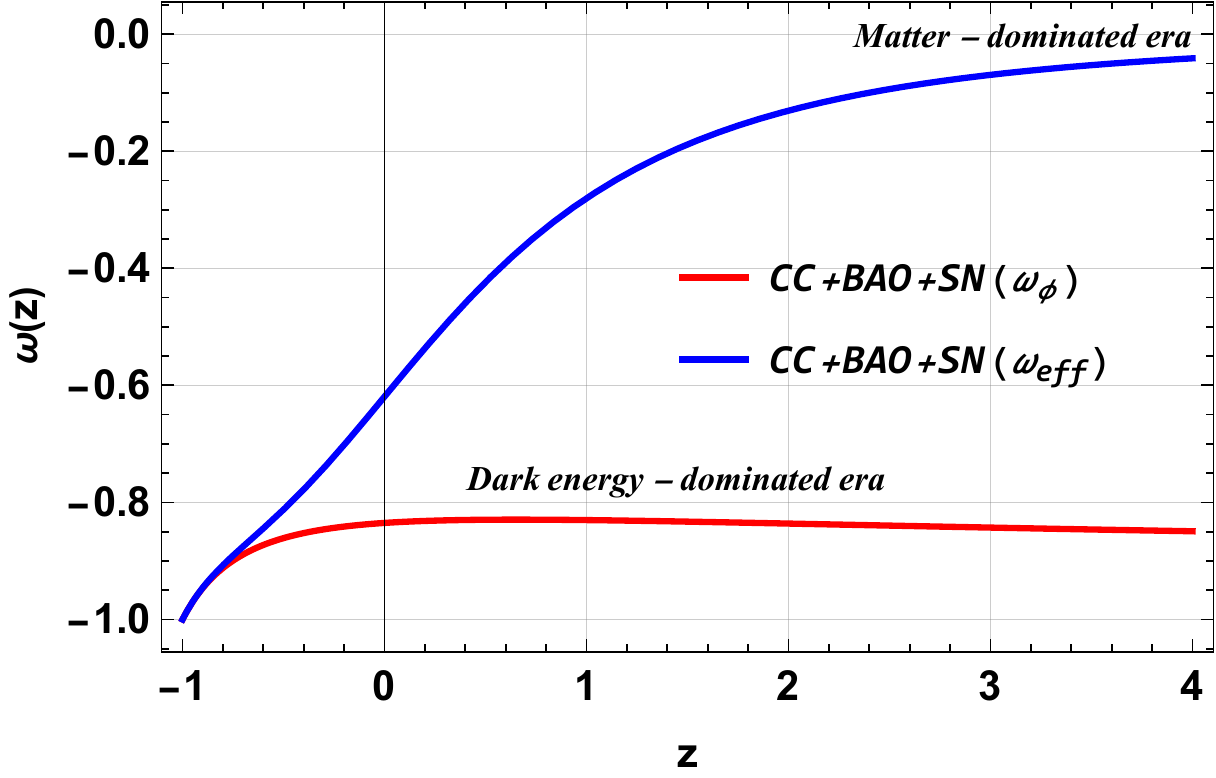}}
\caption{The plot shows the relation between the EoS parameter ($\omega_{\phi}$ and $\omega_{eff}$) and redshift ($z$) using the values constrained from the combined $CC$+$BAO$+$SN$ data-sets.}
\label{F_EoS}
\end{figure}

From Fig. (\ref{F_EoS}), it is seen that both the EoS parameter for the scalar field and the effective EoS parameter exhibit accelerating behavior. The effective EoS parameter begins in the matter-dominated region and progresses through the quintessence phase before reaching a constant value in the cosmological constant region. While the EoS parameter for the scalar field shows the behavior of the quintessence throughout the cosmic evolution as expected and tends to the cosmological constant in the future, which leads to behavior similar to the effective EoS parameter. As mentioned in Sec. \ref{sec1}, the literature extensively discusses the quintessence behavior of DE in the presence of a scalar field source, employing various parameterizations. In our study, we have substantiated this quintessence behavior by determining the current values of the scalar field EoS parameter. The constrained values of the model parameters correspond to $\omega_{0}=-0.84^{+0.06}_{-0.05}$. This result provides strong evidence for the quintessence nature of DE and aligns with previous findings in the field \cite{Hernandez, Zhang}. In previous studies, Singh and Nagpal (2020) explored the behavior of the EoS for DE using the EDSFD parametrization, finding a range of $-1 < \omega_{0} < -0.2$. On the other hand, Debnath and Bamba \cite{Debnath} investigated different parametrizations including Linear, CPL, JBP, and Efstathiou, and obtained specific values for $\omega_{0}$: -0.738, -0.796, -0.755, and -0.765, respectively.

\begin{figure}[h]
\centerline{\includegraphics[scale=0.70]{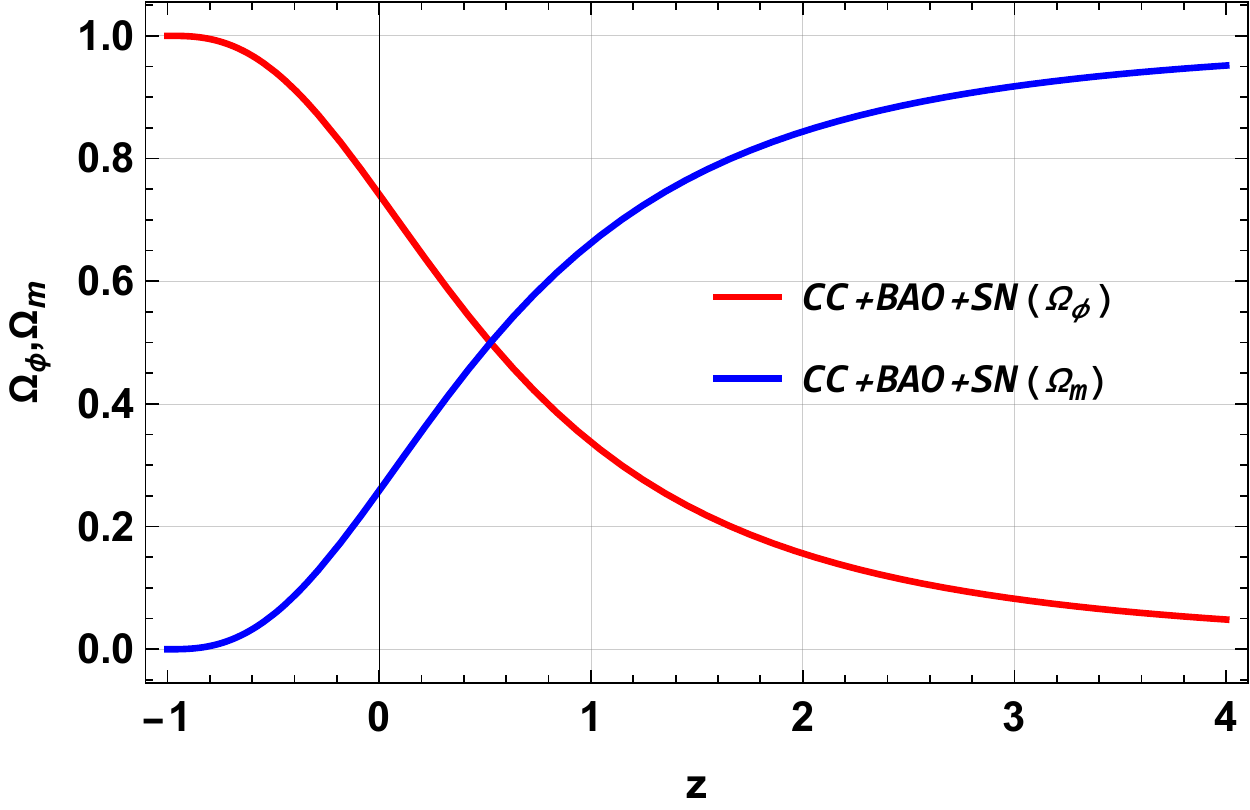}}
\caption{The plot shows the relation between the density parameters ($\Omega_{\phi}$ and $\Omega_{m}$) and redshift ($z$) using the values constrained from the combined $CC$+$BAO$+$SN$ data-sets.}
\label{F_Omega}
\end{figure}

Fig. \ref{F_Omega} depicts the evolution density parameter for matter and scalar field. According to Fig. \ref{F_Omega}, the early Universe is dominated by non-relativistic matter, such as dark matter and baryonic matter, while the scalar field density parameter is negligible. When the Universe expands, the matter density parameter decreases as the volume of the Universe increases, but the scalar field density parameter becomes dominant at later times, leading to an acceleration of the expansion of the Universe. For this model, using the combined $CC$+$BAO$+$SN$ data-sets, we found the best-fit value of the matter density parameter as $\Omega_{m0}=0.25849^{+0.00026}_{-0.00025}$, which is somewhat lower than the value reported by the Planck measurement \cite{Planck2020}.

\begin{figure}[h]
\centerline{\includegraphics[scale=0.72]{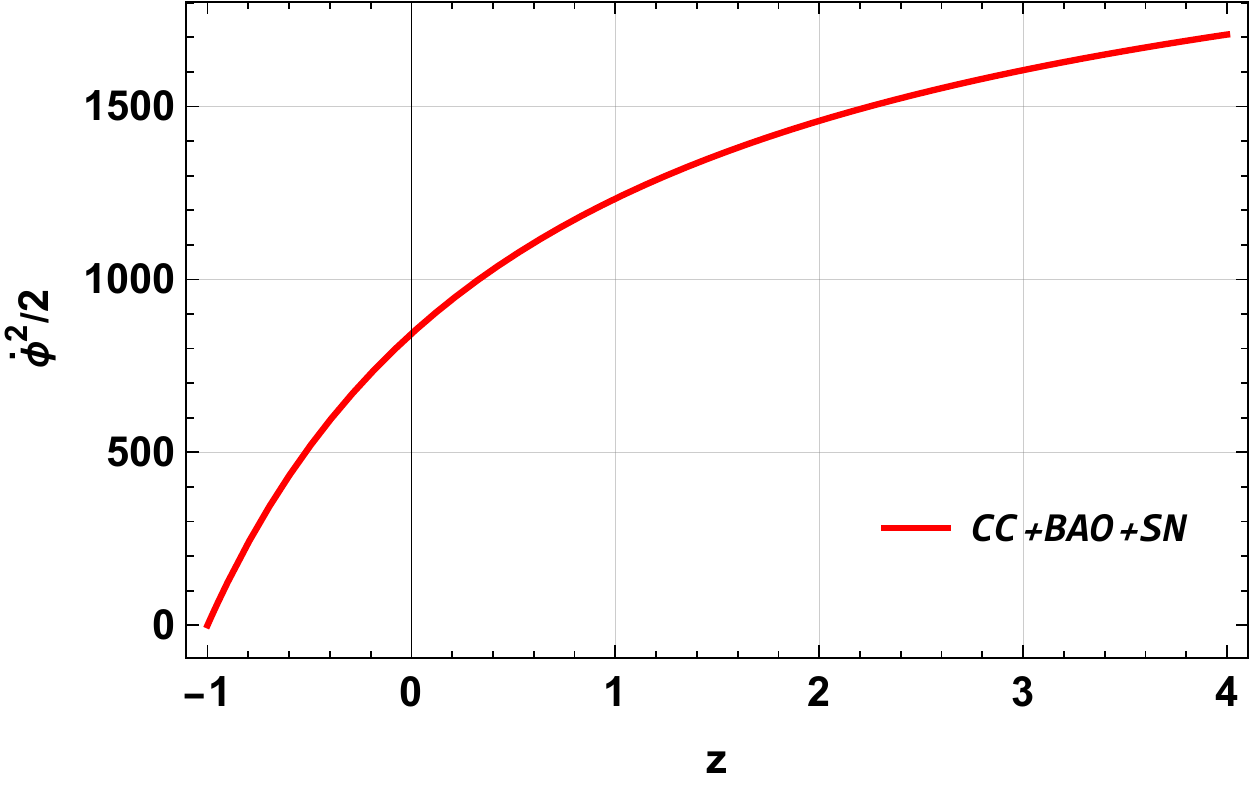}}
\caption{The plot shows the relation between the kinetic energy for the scalar field ($\frac{\overset{.}{\phi }^{2}}{2}$) and redshift ($z$) using the values constrained from the combined $CC$+$BAO$+$SN$ data-sets.}
\label{F_phi}
\end{figure}

\begin{figure}[h]
\centerline{\includegraphics[scale=0.72]{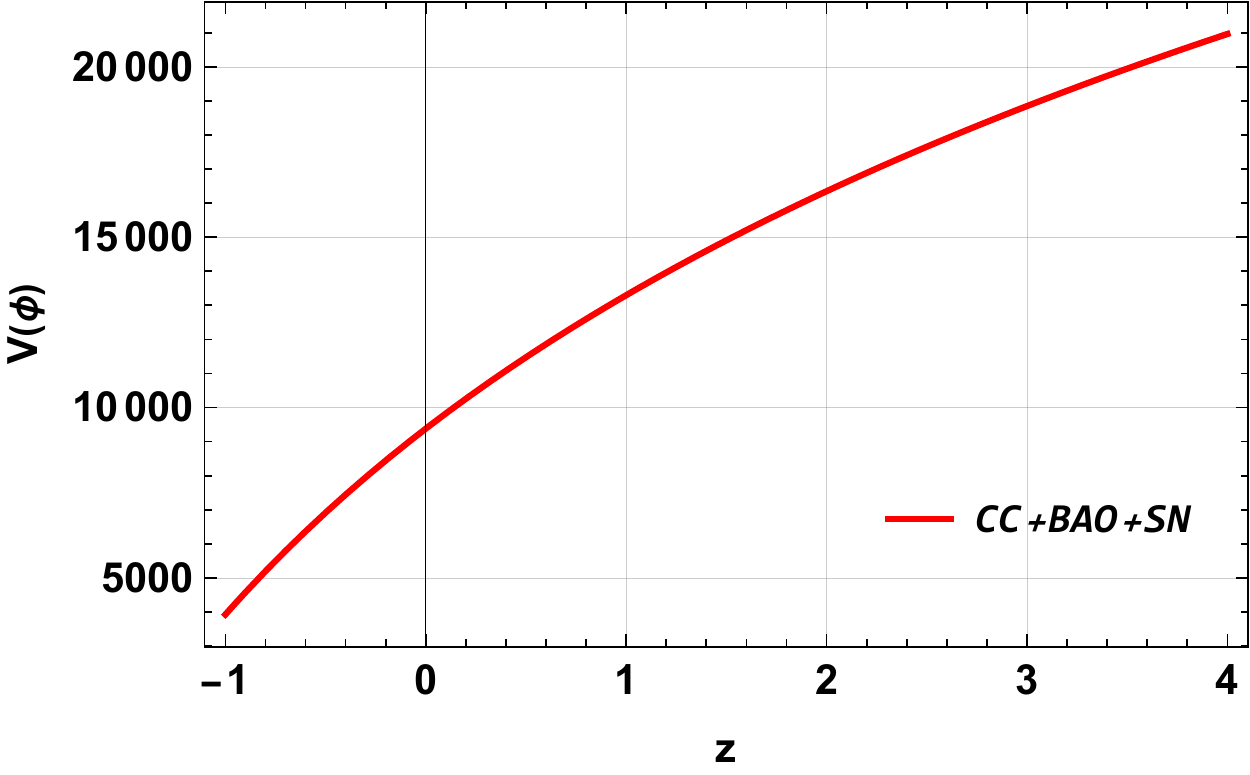}}
\caption{The plot shows the relation between the potential energy for the scalar field ($V(\phi)$) and redshift ($z$) using the values constrained from the combined $CC$+$BAO$+$SN$ data-sets.}
\label{F_V}
\end{figure}

The scalar field, which is responsible for DE, is a mysterious kind of energy that pervades the Universe and is assumed to be driving the Universe's accelerating expansion. Figs. \ref{F_phi} and \ref{F_V} also show the evolution of the kinetic and potential energy of the scalar field. As time passes, the scalar field changes from a high-energy state to a lower-energy one. This can be observed in the behavior of kinetic and potential energy, which both decrease from high positive to low positive values \cite{Pacif,Kar,Sharma,S1,S2}.

\section{Statefinder diagnostic}
\label{sec5}

The study of the geometrical parameters of the Universe is a fundamental part of modern cosmology. The statefinder pair ($r$, $s$), which are geometric quantities directly derived from the metric, is an essential diagnostic tool used to examine the nature of DE. The statefinder parameters are defined by Sahni et al. \cite{Sahni, Alam} as follows:
\begin{equation}
r=\frac{\overset{...}{a}}{aH^{3}}=2q^{2}+q-\frac{\overset{.}{q}}{H},  
\end{equation}%
\begin{equation}
s=\frac{\left( r-1\right) }{3\left( q-\frac{1}{2}\right) }.
\end{equation}

\begin{figure}[h]
\centerline{\includegraphics[scale=0.70]{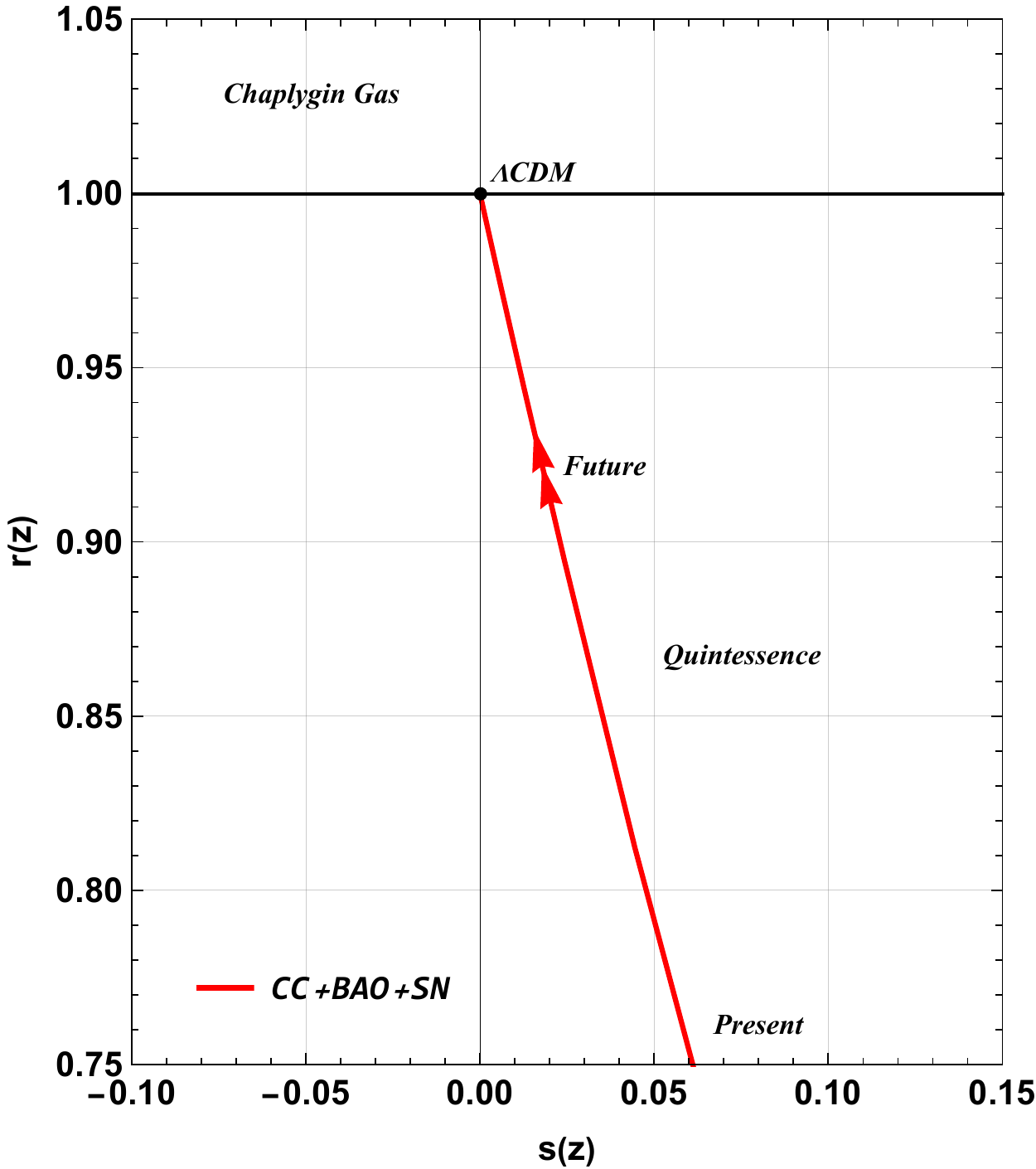}}
\caption{The plot shows the $r-s$ plane using the values constrained from the combined $CC$+$BAO$+$SN$ data-sets with $-1\leq z\leq4$.}
\label{F_rs}
\end{figure}

\begin{figure}[h]
\centerline{\includegraphics[scale=0.70]{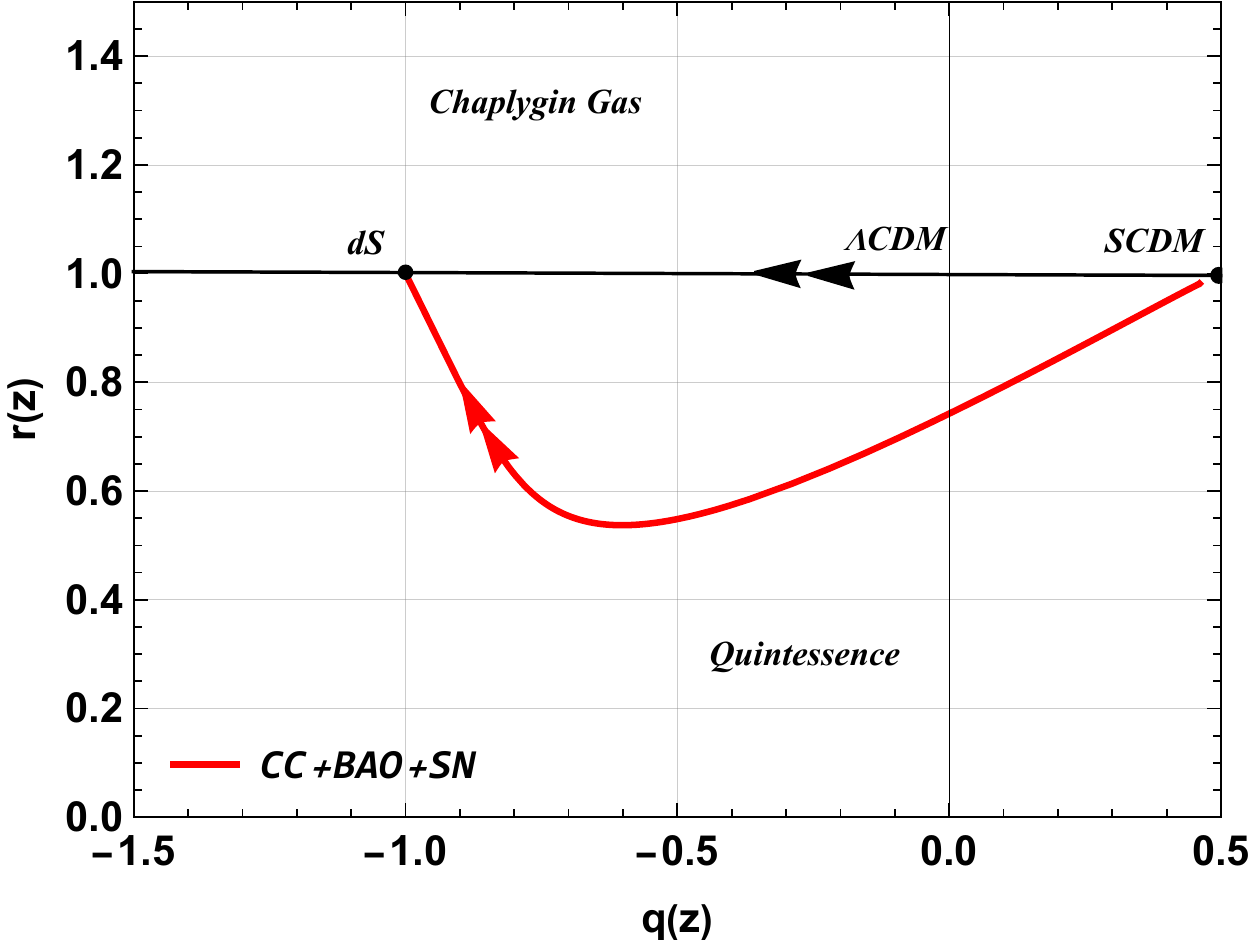}}
\caption{The plot shows the $r-q$ plane using the values constrained from the combined $CC$+$BAO$+$SN$ data-sets with $-1\leq z\leq4$.}
\label{F_rq}
\end{figure}

The dimensionless statefinder parameters $r$ and $s$ are coupled to the higher derivatives of the scale factor. They are model-independent and can differentiate between several DE scenarios based purely on Universe geometry. Especially, for the spatially flat $\Lambda$CDM model, the statefinder parameters are $r=1$ and $s=0$. Different DE models have various values for the statefinder parameters. For example, the quintessence model has $r<1$ and $s>0$. The Chaplygin gas model has $r>1$ and $s<0$. Finally, the holographic DE model has $r=1$ and $s=\frac{2}{3}$. According to Fig. \ref{F_rs} ($r-s$ plane), the model under consideration has starting values of $r<1$ and $s>0$, indicating that the scalar field in the Universe behaves as a quintessence. However, as time progresses, the model approaches the $\Lambda$CDM model with $r=1$ and $s=0$. On the other hand, Fig. \ref{F_rq} ($r-q$ plane) also suggests that while the Universe in the model is currently filled with a quintessential fluid, it is expected to gradually de-Sitter (dS) phase ($r=1$ and $q=-1$), in which the Universe is dominated by a cosmological constant, where the expansion of the Universe is accelerating at a constant rate. We can conclude the behavior of the statefinder parameters in the model being studied is consistent with the behavior of the cosmological parameters discussed in the previous section. 

\section{Accretion process in scalar field DE model}
\label{sec6}

The study of black hole accretion is an important topic of research in astrophysics. As matter falls into a black hole, it is heated to extremely high temperatures and emits radiation that astronomers can observe. To investigate the accretion process, it is frequently believed that the black hole fluid represents a tiny fraction of the Universe's total matter content and that no matter from the black hole is transformed from dark matter to DE as the Universe expands. This assumption is based on the premise that black holes emerge as a result of the collapse of massive stars, which are predominantly constituted of baryonic matter. On the other hand, dark matter is assumed to be a non-baryonic kind of matter that interacts with other forms of matter, including black holes, very weakly. As a result, it is widely considered that the amount of dark matter falling onto a black hole during the accretion process is insignificant in comparison to the amount of baryonic matter. In this paper, we consider the accretion equation developed by Babichev et al. \cite{Babichev}, which is premised on the conservation of the energy-momentum tensor of a perfect non-self-gravitating fluid in the Schwarzschild space-time \cite{Michel}, and a mass variation term that can be supported by geometrical properties of the energy-momentum tensor in diagonal metrics \cite{Lima}.

For an asymptotic observer, the black hole mass rate can be written as \cite{Babichev,Lima,Martin},
\begin{equation}
\overset{.}{M}=4\pi AM^{2}\left( \rho _{eff}+p_{eff}\right), 
\label{BH}
\end{equation}
where $A$ is a constant \cite{Babichev}. $M$ represents the mass of the black hole and $\rho _{eff}=\rho _{m}+\rho _{\phi }$ and $p_{eff}=p_{\phi }$ are the total (effective) energy density and pressure of the Universe.

Using Eqs. (\ref{F1}), (\ref{F2}) and (\ref{H}) in Eq. (\ref{BH}), we obtain the black hole mass as a function of redshift,
\begin{equation}
M(z)=\frac{1}{8 \pi  A H_{0} \sqrt{\Omega
_{m0} (z+1)^3+\log (\alpha +\beta  z)}-c_1},
\end{equation}
where $c_1$ is a constant of integration.

\begin{figure}[h]
\centerline{\includegraphics[scale=0.70]{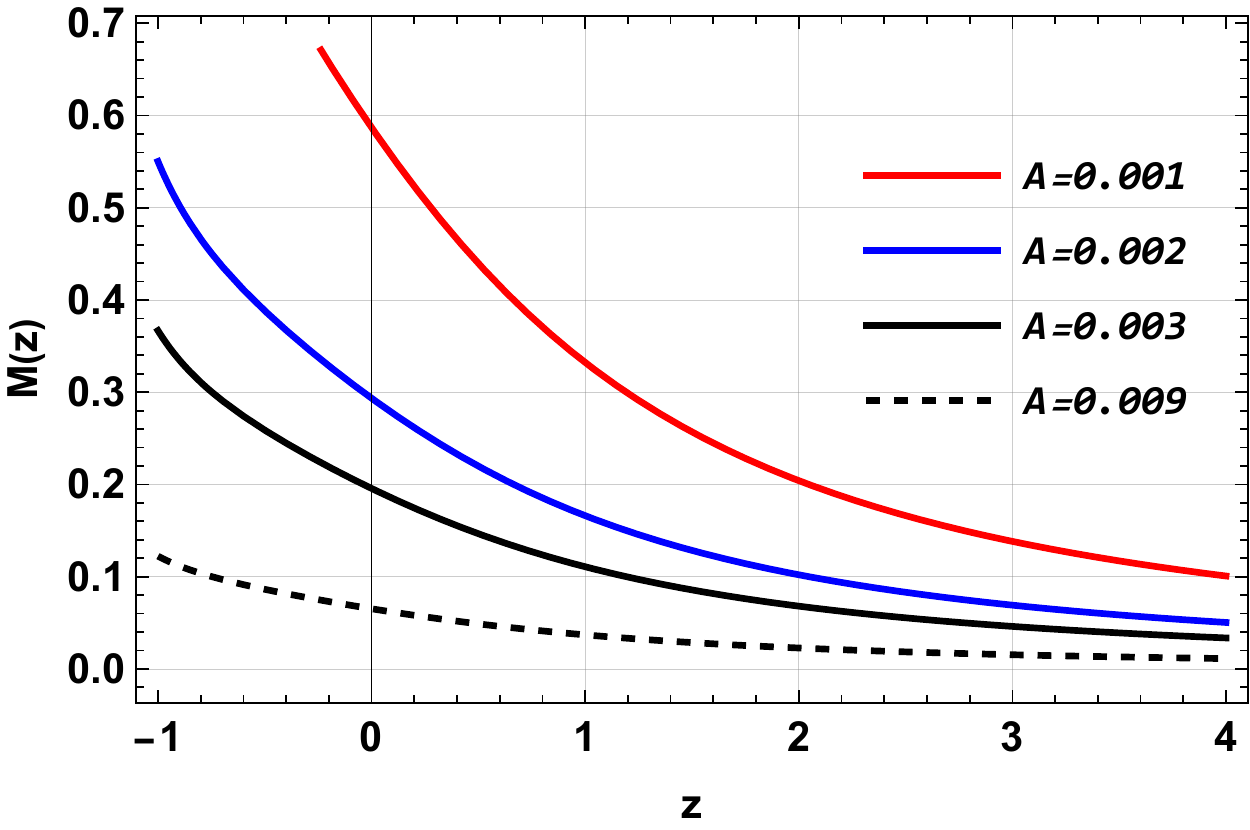}}
\caption{The plot shows the relationship between the black hole mass ($M(z)$) and redshift ($z$) using the values constrained from the combined $CC$+$BAO$+$SN$ data-sets and $c_{1}=0$.}
\label{F_M}
\end{figure}

As seen in Fig. \ref{F_rho}, the matter gets diluted quicker than DE because a huge quantity of energy transfers from matter to DE as the Universe expands. Fig. \ref{F_M} shows that the mass of black holes in this scenario increases over time for some values of $A$. Fig. \ref{F_M} shows that the mass of black holes in this scenario increases over time for some values of $A$. The rate of growth decreases and finally ceases when DE becomes the dominant component of the Universe. Furthermore, the figure implies that in this scenario, black holes can reach a maximum mass. This is due to the fact that as the Universe expands, the available mass for black holes to accrete diminishes, eventually leading to a point when black holes can no longer accumulate mass. These findings are consistent with the predictions by Lima et al. \cite{Lima} for the $\Lambda$CDM model.

\section{Conclusion}
\label{sec7}

The concept of DE in the Universe is one of the most exciting and difficult questions in modern cosmology. DE is a mysterious type of energy that appears to pervade the Universe and is assumed to be responsible for the Universe's accelerating expansion. Despite years of research and observation, the nature of DE is still completely unknown, posing a tremendous challenge to our understanding of the Universe. In this paper, we developed an FLRW cosmology model with an acceptable parametrization for scalar field energy density as a logarithmic function of redshift in the framework of the standard theory of gravity, which supports the necessary transition from the decelerated to the accelerated periods of the Universe. We have obtained an exact solution of Einstein’s field equations with a scalar field source of DE, which consists of three model parameters. Moreover, utilizing a combination of CC data-sets, BAO, and recently published Pantheon data-sets, we determined the best-fit values for the model parameters. The resulting best-fit values are $H_{0}=67.79_{-0.59}^{+0.59}$ $km/s/Mpc$, $\Omega_{m0}=0.25849^{+0.00026}_{-0.00025}$, and $\beta=0.77^{+0.24}_{-0.23}$ for the combined $CC$+$BAO$+$SN$ data-sets (Tab. \ref{tab}). Our findings for $H_0$ (Fig. \ref{ErrorHubble}) and $\Omega_{m0}$ (Fig. \ref{F_Omega}) are highly consistent with contemporary measurements, which were derived from the Planck satellite's observations and estimated as $H_0=67.4\pm0.5$ $km/s/Mpc$ and $\Omega _{m}^{0}=0.315\pm 0.007$ \cite{Planck2020}. Furthermore, our results are in agreement with other research studies that have employed similar techniques to estimate the values of both $H_0$ and $\Omega_{m0}$ \cite{Chen1, Chen2, Aubourg, Capozziello}.

Furthermore, we examined the dynamics of the deceleration parameter and the densities of both matter and scalar field DE for the constrained values of the model parameters. Fig. \ref{F_q} illustrates the evolution of the deceleration parameter, indicating a recent transition of the Universe from a decelerated to an accelerated phase. The transition redshift is $z_{tr}=0.79^{+0.02}_{-0.02}$ corresponding to the combined $CC$+$BAO$+$SN$ data-sets. The present value of the deceleration parameter is $q_{0}=-0.43^{+0.06}_{-0.06}$. The energy densities presented in Fig. \ref{F_rho} display a positive behavior as anticipated. Fig. \ref{F_EoS} demonstrates the evolution of the EoS parameter, indicating that our cosmological model adheres to the quintessence scenario with the present value $\omega_{0}=-0.84^{+0.06}_{-0.05}$. We have also discussed the behavior of kinetic energy and potential energy of the scalar field in Figs. \ref{F_phi} and \ref{F_V}. Figs. \ref{F_rs} and \ref{F_rq} illustrate the evolution of the statefinder diagnostic, suggesting a quintessence behavior, which is in concurrence with the other cosmological parameters.

Lastly, our investigation also involved examining the evolution of black holes in the presence of both matter and DE, with consideration given to the Schwarzschild-type metric in the vicinity of the black hole. Fig. \ref{F_M} displays the representation of the black hole mass as a function of redshift, which was derived by determining the energy densities of both the matter and scalar field DE components. It is worth noting that the black hole fluid constitutes an insignificant fraction of the overall matter content. Our findings suggest that black holes within this scenario exhibit an initial increase in mass. However, at later epochs, the growth in mass ceases as a result of DE accretion. Moreover, as cosmic time increases substantially, the mass of the black holes reaches a maximum value \cite{Lima}. Therefore, we can confidently say that our model provides a viable explanation for the observed phenomena in the Universe. By accurately representing the observed data, our model supports the current understanding of the Universe and provides a foundation for further inquiry. The success of our model in closely matching it with the observed data is an important step forward in our efforts to understand the evolution of the Universe.

\section*{Acknowledgments}
This research project is sponsored by QingLan Project (SG050402010423) and Key Program of Natural Science of Changzhou College of Information Technology (CXZK202203Y). M. Koussour is thankful to Dr. Shibesh Kumar Jas Pacif, Centre for Cosmology and Science Popularization, SGT University for some useful discussions. N. Myrzakulov is thankful to the Science Committee of the Ministry of Science and Higher Education of the Republic of Kazakhstan (Grant No. AP09058240). G. Mustafa is also very thankful to Prof. Xianlong Gao from the Department of Physics, Zhejiang Normal University, for his kind support and help during this research. Further, G. Mustafa acknowledges Grant No. ZC304022919 to support his Postdoctoral Fellowship at Zhejiang Normal University, P. R. China. 

\textbf{Data availability} There are no new data associated with this
article.

\end{document}